\UseRawInputEncoding 
\documentclass[aps,preprint,preprintnumbers,amsmath,amssymb]{revtex4}
\usepackage{amsmath,mathrsfs,amsbsy,color,graphicx,bm,amsthm,amsfonts}
\usepackage{units}
\usepackage{bbm}
\usepackage{times}
\usepackage{dcolumn}
\usepackage{mathrsfs}
\usepackage{amsmath,amssymb,epsfig}
\usepackage{hyperref}
\begin{document}

\title{Distribution  and generation of quantum coherence for Gaussian states in de Sitter space }
\author{Qianqian Liu$^{1}$, Cuihong Wen$^{1}$ \footnote{Email: cuihong\_wen@hunnu.edu.cn}, Jieci Wang$^{1}$\footnote{Email: jcwang@hunnu.edu.cn},  Jiliang Jing$^{1}$\footnote{Email: jljing@hunnu.edu.cn}}
\affiliation{$^1$ Department of Physics, and Synergetic Innovation Center for Quantum Effects \\
and Applications,Hunan Normal University, Changsha, Hunan 410081, China}


\begin{abstract}

We study the distribution and generation  of quantum coherence for two-mode and multi-mode  Gaussian states  in de Sitter space.
It is found that the quantum coherence  is redistributed among  the mode in different open charts under the curvature effect of de Sitter space.   In particular,  the Gaussian coherence for the  initially  correlated  state   is found to survive in the limit of infinite curvature, while quantum entanglement vanishing in this limit. Unlike  entanglement  and steering,  the coherence of a massive scalar  field is more robust than  a massless  field under the influence of curvature of  de Sitter space. In addition, it is shown that the curvature   generates two-mode Gaussian state and three-mode Gaussian state quantum coherence among the open charts, even though the observers are localized in causally disconnected regions. It is worth noting that the gravity-generated three-mode coherence is extremely sensitive to the curvature effect for the conformal and massless scalar fields, which  may be in principle employed  to  design an effective detector for the space  curvature.
\end{abstract}

\vspace*{0.5cm}

\maketitle
\section{Introduction}

The Quantum state superposition principle is one of  the most fundamental features of quantum  theory,  which distinguishes classical  theory.   As a key aspect of quantum physics, quantum coherence is an embodiment of the superposition principle of states and  is the basis of phenomena such as quantum entanglement and multi-particle interference \cite{C,coherencer,coherencer1}. With the help of quantum coherence, one can implement various quantum information processing tasks that cannot
be accomplished classically, such as quantum computing  \cite{computing1, computing2}, quantum metrology \cite{metrology1, metrology2}, quantum biology \cite{biolo1,photosynthetic}, and other
 quantum  information storage \cite{trans2} and transmission \cite{trans1,Tele} tasks. Despite the fundamental importance of quantum coherence, the quantum resource study of  coherence  receives
increasing attention until Baumgratz \emph{et. al.} introduced the framework for the measurement of
coherence \cite{C}. Motivated by this, more and more measurements of coherence have been performed  \cite{C1,C2,C3,C4,Grover} in recent years.  Recently,
Xu proposed a quantification of coherence  for continuous-variable  quantum systems \cite{XU},  which performs the framework of a quantum  resource theory for infinite-dimensional quantum states in quantum optics   \cite{inform1}.

Understanding quantum effects in the framework of relativity \cite{relat2,mode,RQI1,relat1,RQI2,Frris,RQI4,RQI5,relat4,RQI6,stee,Gyong1,Gyong2,DW1,DW2,DW3} is essential.
 For example, it was found that  nonclassical correlations are generated between the open charts  in the exponentially expanding de Sitter space\cite{DS1,DS,DS2,DS3,DS4,DS5,tripartite}.  Interestingly, the entanglement remains nonzero even if the distance between the two regions becomes larger
than the Hubble length. This means there exists a
nonclassical correlation between two causally disconnected regions
and the existence of the entanglement means that the Reeh-Schlieder theorem \cite{DS6} holds in de Sitter space. People cannot understand this phenomenon without the combination of quantum information theory and  relativity.  In addition, the  quantum discord \cite{discord}  and quantum steering \cite{steeringw} of  field modes  are also found to be generated between two disconnected open charts  under the  influence of the space curvature of  de Sitter space.

In this paper, we study the distribution and generation of  continuous-variable quantum  coherence in the background of de Sitter space.
 The considered initial state is a two-mode squeezed Gaussian state, which can be employed  to define particle states when the space-time has at least two asymptotically flat regions \cite{mode,RQI2,RQI4,RQI5,tripartite}.  It has a special role in quantum field
theory because the  field modes in causally disconnected regions  are found to be pair-wise squeezed
under the influence of spacetime structure evolution and relativistic effects \cite{relat2, mode, RQI1,
tripartite,steeringw}. In addition,  as  the most typical  entangled state for continuous variables,  the  two-mode squeezed state can
be produced with existing technology  and be exploited for a few important  continuous-variable
 quantum information tasks \cite{brareview}.
 It is found that  two-mode and three-mode quantum coherence are generated among the open charts, even though the observers are causally disconnected.

The organization of the paper is as follows. In Sec. II we review the dynamics of field modes in de Sitter space and through the Bogoliubov transformations, the Bunch-Davies vacuum can be represented by the open charts vacuum of the de Sitter space. In Sec. III we discuss the method of measuring the quantum coherence of continuous variables. In Sec. IV we study the behavior of quantum coherence between the two-mode and three-mode Gaussian states in de Sitter space. In the final section, we summarize our results.

\section{Quantization of  scalar field in de Sitter space \label{model}}

In this section, we review the method developed by Maldacena and Pimentel to obtain the quantization of the scalar field in de Sitter space \cite{DS2,DS1}  with some comments on the parameters that relevant to the discussion of quantum  coherence.
The four-dimensional Euclidean de Sitter space can be embedded in the Five-dimensional Euclidean space. Therefore,
the coordinate frames of open charts in de Sitter space can be obtained by analytic continuation from the Euclidean metric \cite{DS} respectively,
\begin{eqnarray}\label{eq1}
ds^2_R&=&H^{-2}\left[-dt^2_R+\sinh^2t_R\left(dr^2_R+\sinh^2r_R\,d\Omega^2\right)
\right]\,,\nonumber\\
ds^2_L&=&H^{-2}\left[-dt^2_L+\sinh^2t_L\left(dr^2_L+\sinh^2r_L\,d\Omega^2\right)
\right]\,,
\end{eqnarray}
where $H^{-1}$ is the Hubble radius and $d\Omega^{2}$ is the metric on the two-sphere. 

The solutions of the Klein-Gordon equation  in de Sitter space are found to be $u_{\sigma plm}(t,r,\Omega)\sim\frac{H}{\sinh t_{R(L)}}\chi_{p,\sigma}(t_{R(L)})\,Y_{p\ell m} (r_{R(L)},\Omega)$,
where $Y_{p\ell m}$ is a harmonic function on the three-dimensional hyperbolic space. The positive frequency mode functions $\chi_{p,\sigma}(t_{R(L)})$  corresponding to Euclidean vacuum(the Bunch-Davies vacuum), which are supported in both $R$ and $L$ regions derived
by Sasaki, Tanaka and Yamamoto in \cite{DS1}.
The index $\sigma=\pm1$ distinguishing the independent solutions for each open charts and
the parameter $p$ is regarded as the curvature parameter of de Sitter space because the effect of curvature appears around $p\sim1$ and  increases through $p$ less than 1.  In addition, the limit of infinite curvature appears at $p\rightarrow0$ \cite{discord,tripartite}. The normalization coefficient for function $\chi_{p,\sigma}(t_{R(L)})$ is
 $N_{p}=\frac{4\sinh\pi p\,\sqrt{\cosh\pi p-\sigma\sin\pi\nu}}{\sqrt{\pi}\,|\Gamma(\nu+ip+\frac{1}{2})|}\, $. The  mass parameter $ \nu$ is defined by  $ \nu=\sqrt{\frac{9}{4}-\frac{m^2}{H^2}}\, $, which has two special values:
 $\nu=1/2$  for the conformally coupled
 scalar field, and $\nu=3/2$ for the minimally coupled massless  scalar field.

With these positive frequency mode functions, the scalar field can be expanded in terms of the creation and annihilation operators
\begin{eqnarray}\label{eq4}
\hat\phi(t,r,\Omega)
=\frac{H}{\sinh t}\int dp \sum_{\sigma,\ell,m}\left[\,a_{\sigma p\ell m}\,\chi_{p,\sigma}(t)
+a_{\sigma p\ell -m}^\dagger\,\chi^*_{p,\sigma}(t)\right]Y_{p\ell m}(r,\Omega)
\,,
\end{eqnarray}
where $a_{\sigma p\ell m}|0\rangle_{\rm BD}=0$ and the commutation relations are $[a_{\sigma p\ell m},a^{\dag}_{\sigma^{\prime} p^{\prime}\ell^{\prime} m^{\prime}}]=\delta(p-p^{\prime})\times\delta_{\sigma,\sigma^{\prime}}\delta_{l,l^{\prime}}\delta_{m,m^{\prime}}$.  In the following the indices $p,l,m$ of the operators and mode functions
are omitted for simplicity.

By  introducing the  creation and annihilation operators in different open charts, one can calculate  the Bogoliubov transformations  \cite{DS1,DS2} between the operators which  are defined in the Bunch-Davies vacuum and  the open charts vacua, respectively. Then the Bunch-Davies vacuum is found to be   \cite{DS1,DS2}
\begin{eqnarray}\label{gammap3}
|0\rangle_{\rm BD}=\sqrt{1-|\gamma_p|^{2}}\sum_{n=0}^{\infty}(\gamma_p)^{n}|n\rangle_{\rm L}|n\rangle_{\rm R},
\end{eqnarray}
where we defined $|n;plm\rangle=\frac{1}{\sqrt{n!}}(c^{\dag}_{R})^{n}|0\rangle_{R^{\prime}}$, and $|\gamma_p|<1$ should be satisfied. In Eq. (\ref{gammap3}), the parameter $\gamma_p$ is
\begin{eqnarray}\label{gammap2}
\gamma_p = i\frac{\sqrt{2}}{\sqrt{\cosh 2\pi p + \cos 2\pi \nu}
 + \sqrt{\cosh 2\pi p + \cos 2\pi \nu +2 }}\,.
\end{eqnarray}
For the conformally coupled  scalar field ($\nu=1/2$) and the minimally coupled massless scalar field ($\nu=3/2$),  it  simplifies to $|\gamma_p|=e^{-\pi p}$.

\section{Measurement of quantum coherence for continuous variables \label{GSteering}}
For a two-mode  continuous variable quantum  system, we
 define the vector of field quadratures (âpositionâ and âmomentumâ) operators as $\hat{R}=(\hat{x}_{A},\hat{p}_{A},\hat{x}_{B},\hat{p}_{B})$, which are related to the annihilation $\hat{a}_{i}$ and creation  $\hat{a}_{i}^{\dag}$
operators of each mode, by the relations $\hat{x}_{i}=\frac{(\hat{a}_{i}+\hat{a}_{i}^{\dag})}{\sqrt{2}}$
and $\hat{p}_{i}=\frac{(\hat{a}_{i}-\hat{a}_{i}^{\dag})}{\sqrt{2}i}$. The vector operator satisfies the commutation relationship: $[{{{\hat R}_i},{{\hat R}_j}} ] = i{\Omega _{ij}}$, with $\Omega  =  \bigoplus_1^{n+m} {{\ 0\ \ 1}\choose{-1\ 0}}$ being  symplectic form.
The first and second moments of a two-mode Gaussian state ${\rho _{AB}}$ can completely describe all its properties. For the bipartite state $\rho_{AB}$, its covariance matrix has the form
\begin{equation}\label{matric1}
\boldsymbol{\sigma}\equiv\left(\begin{array}{cc}
\boldsymbol{\alpha}&\boldsymbol{\gamma}\\
\boldsymbol{\gamma}^{T}&\boldsymbol{\beta}
\end{array}\right)\, ,
\end{equation}
 where $\alpha = {\rm dig}(a,a)$, $\beta = {\rm dig}(b,b)$, and $\gamma={\rm dig}(c_{+},c_{-})$  are $2\times2$ real matrices. The   symplectic eigenvalues of the two-mode Gaussian covariance matrix Eq. (\ref{matric1}) are $2{\nu}_{\mp}^2=\Delta \mp\sqrt{\Delta^2-4I_4}$ with $\Delta=a^2+b^2+2c_{+}c_{-}$, $I_4=det(\sigma)$.

As shown in \cite{XU}, the continuous variable quantum coherence of a $m$-mode Gaussian states can be measured by  $C(\rho)=\inf{S(\rho ||\delta)}$, where $S(\rho||\delta)= tr(\rho \log_{2}\rho) - tr(\rho \log_{2}\delta)$ is the relative entropy, $\delta$ is an incoherent Gaussian state  and $inf$ runs over all incoherent Gaussian states. In addition, The entropy of $\rho$ is defined by \cite{entropy}
 \begin{eqnarray}\label{coherence1}
 S(\rho)=-tr(\rho \log_{2}\rho)=\sum_{i=1}^{m}f({\nu}_i),
 \end{eqnarray}
where $f({\nu}_i)=\frac{{\nu}_i+1}{2}\log_2\frac{{\nu}_i+1}{2}-\frac{{\nu}_i-1}{2}\log_2\frac{{\nu}_i-1}{2}$, and $\nu_i$ is symplectic eigenvalues of each mode.
 Then one obtains the definition of the quantum coherence of Gaussian states \cite{XU}
\begin{eqnarray}\label{coherence3}
\nonumber C({\rho})=&-&S(\rho)+\sum_{i=1}^{m}[(\overline{n}_i+1)\log_2(\overline{n}_i+1)\\
&-&\overline{n}_i\log_2\overline{n}_i],
\end{eqnarray}
where $\overline{n}_i=\frac{1}{4}(\sigma_{11}^i+\sigma_{22}^i+[d_1^i]^2+[d_2^i]^2-2)$ is the mean occupation value.
Here, $\sigma^i$ are elements of the subsystem of mode $i$ in a continuous variable matrix, respectively, and $[d^i]^2$ is $i$ first statistical moment of the $k$ mode.

\section{Distribution of Gaussian quantum coherence in de Sitter space \label{tools}}
\subsection{Reduction of  quantum coherence between initially correlated modes }
In the subsection, we study the Gaussian quantum coherence between the global observer Alice and the open chart observer Bob (in region $R$ ).
 We use the form of the covariance matrix to describe the initial state $\rho_{AB}$ prepared by the two-mode squeezed Gaussian state in Bunch-Davies vacuum
\begin{eqnarray}\label{matric3}
\sigma^{(G)}_{AB}(s)= \left(\!\!\begin{array}{cccc}
\cosh(2s)I_{1}&\sinh(2s)Z_{1}\\
\sinh(2s)Z_{1}&\cosh(2s)I_{1}
\end{array}\!\!\right),
\end{eqnarray}
where $s$ is the squeezing of the initial state and $I_{1}= {{\ 1\ \ 0}\choose{\ 0\ \ 1}},$ $Z_{1}= {{\ 1\ \ 0}\choose{\ 0 \  -1}}$.

From Eq. (\ref{gammap3}), we can see  the Bunch-Davies vacuum for a  global  observer can be expressed as a two-mode squeezed state of the $R$ and $L$ vacua \cite{DS1,DS,DS2,discord,tripartite}.  In the phase space, The two-mode squeezing transformation can be expressed by a symplectic operator \cite{tripartite}
\begin{eqnarray}\label{matric4}
S_{B,\bar B}(\gamma_p)= \frac{1}{\sqrt{1-|\gamma_p|^2}}\left(\!\!\begin{array}{cccc}
I_{1}&|\gamma_p|Z_{1}\\
|\gamma_p|Z_{1}&I_{1}
\end{array}\!\!\right).
\end{eqnarray}

Under this symplectic matrix transformation, the virtual observer anti-Bob is mapped from Bob in the chart $R$ to chart $L$.
Then we can calculate  the  covariance matrix $\sigma^{\rm }_{AB \bar B}$  of the  entire state, which  is
\begin{eqnarray}\label{All3}
\nonumber\sigma^{\rm }_{AB \bar B}(s,\gamma_p) &=& \big[I_A \oplus  S_{B,\bar B}(\gamma_p)\big] \big[\sigma^{\rm (G)}_{AB}(s) \oplus I_{\bar B}\big]\\&& \nonumber\big[I_A \oplus  S_{B,\bar B}(\gamma_p)\big]\\
 &=& \left(
       \begin{array}{ccc}
          \mathcal{\sigma}_{A} & \mathcal{E}_{AB} & \mathcal{E}_{A\bar B} \\
         \mathcal{E}^{\sf T}_{AB} &  \mathcal{\sigma}_{B} & \mathcal{E}_{B\bar B} \\
         \mathcal{E}^{\sf T}_{A\bar B} & \mathcal{E}^{\sf T}_{B\bar B} &  \mathcal{\sigma}_{\bar B} \\
       \end{array}
     \right)
 \,,
\end{eqnarray}
where $\sigma^{\rm (G)}_{AB}(s) \oplus I_{\bar B}$ is the initial covariance matrix for the entire system.
In the above expression the diagonal elements are:  \begin{equation} \mathcal{\sigma}_{A}=\cosh(2s)I_1,\end{equation}
 \begin{equation}
\mathcal{\sigma}_{B}=\frac{|\gamma_p|^2+\cosh(2s)}{1-|\gamma_p|^2}I_1,
\end{equation} and
\begin{equation}\mathcal{\sigma}_{\bar B}=\frac{1+|\gamma_p|^2\cosh(2s)}{1-|\gamma_p|^2}I_1, \end{equation} Similarly,  the non-diagonal elements are found to be
$\mathcal{E}_{AB}=\frac{\sinh(2s)}{\sqrt{1-|\gamma_p|^2}}Z_{1}$, $\mathcal{E}_{B\bar B}=\frac{|\gamma_p|(\cosh(2s)+1)}{1-|\gamma_p|^2}Z_{1}$, and $\mathcal{E}_{A\bar B}=\frac{|\gamma_p|\sinh(2s)}{\sqrt{1-|\gamma_p|^2}}I_{1}$.

Because the charts $R$ and $L$ are causally disconnected  in de Sitter space, Bob can not approach the mode $\bar{B}$ on the other side  of the event horizon.  Therefore, we obtain
the covariance matrix $\sigma_{AB}(s,\gamma_p)$ for Alice and Bob by tracing  over the mode $\bar{B}$
\begin{eqnarray}\label{CMAB}
\sigma_{AB}(s,\gamma_p)= \left(\!\!\begin{array}{cccc}
\cosh(2s)I_{1}&\frac{\sinh(2s)}{\sqrt{1-|\gamma_p|^2}}Z_{1}\\
\frac{\sinh(2s)}{\sqrt{1-|\gamma_p|^2}}Z_{1}&\frac{|\gamma_p|^2+\cosh(2s)}{1-|\gamma_p|^2}I_{1}
\end{array}\!\!\right).
\end{eqnarray}

 We know that the symplectic matrix $\sigma_{AB}$ is a case of the smallest mixed Gaussian state according to the eigenvalues of the partially transposed symplectic matrix.
 For Eq. (\ref{CMAB}), we obtain $\Delta^{(AB)}=1+\frac{(1+|\gamma_p|^2\cosh(2s))^2}{(1-|\gamma_p|^2)^2}$, and $I^{AB}_{4}=\frac{(1+|\gamma_p|^2\cosh(2s))^2}{(1-|\gamma_p|^2)^2}$.
This mean occupation numbers operator for each mode from the covariance matrix is
\begin{eqnarray}\label{nAB}
   \bar{n}_{A}=\frac{1}{2}(\cosh(2s)-1),   \ \nonumber\\ \bar{n}_{B}=\frac{1}{2}(\frac{|\gamma_p|^2+\cosh(2s)}{1-|\gamma_p|^2}-1),
\end{eqnarray}
The quantum coherence of this state can be calculated via Eq. (\ref{coherence3}).

\begin{figure}[htbp]
\centering
\includegraphics[height=2.2in,width=2.8in]{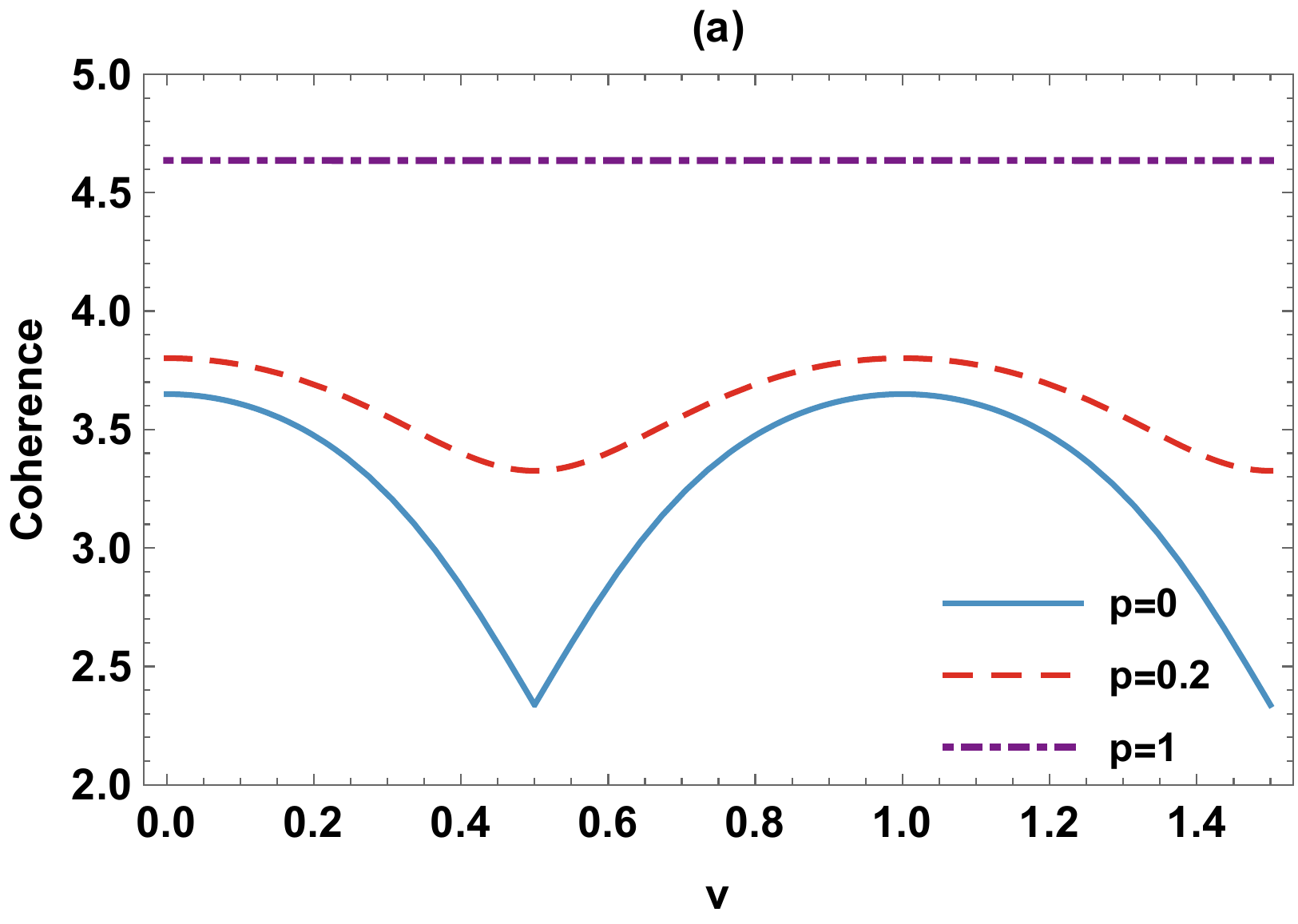}
\includegraphics[height=2.2in,width=2.8in]{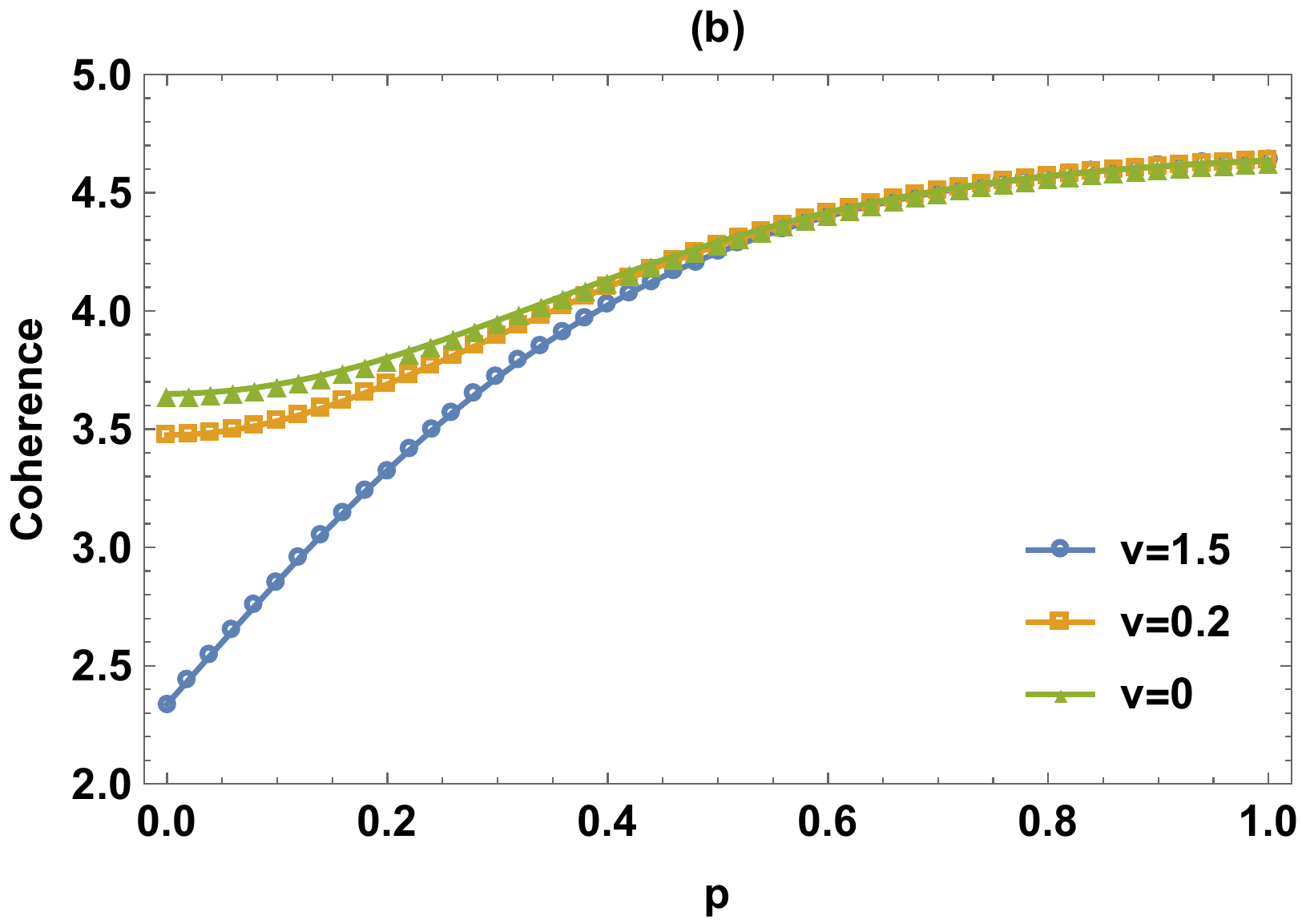}
\caption{ (Color online). (a) Plots of the quantum  coherence as a function of the mass parameter $\nu$, where $p=1$ (purple dotted line), $p=0$ (blue solid line), $p=0.2$ (red dotted line). (b) The  Gaussian quantum coherence between Alice and Bob as a function of the curvature parameter $p$. The green triangle line is for $\nu=0$,
the orange rectangular line is for $\nu=1/5$ and the blue circular line  is for $\nu = 3/2$. Fix the initial state squeezing parameter $s=1$.  }\label{Fig1}
\end{figure}
In Fig. 1(a) and Fig. 1(b), we plot the accessible quantum coherence between Alice and Bob as functions of  the curvature parameter $p$ and mass parameter $\nu$ for a fixed initial  squeezing parameter $s=1$.
 Fig. 1(a) illustrates that the Gaussian  quantum  coherence is not influenced by the mass parameter $\nu$ in the flat space limit $p=1$. For any $p\neq1$, the degree of quantum  coherence  is a periodic function of  the mass parameter $\nu$  with period 1. In the infinite curvature limit $p=0$, the Gaussian quantum coherence reaches { \it non-zero} minimum values for  $\nu=1/2$ (conformal scalar limit) and $\nu=3/2$ (massless scalar limit).
 It is a worthing note that the  behavior of quantum coherence for the Gaussian state is very different than that of entanglement because the  Gaussian  entanglement for the state $\rho_{AB}$ directly reduces to zero in the infinite curvature limit \cite{tripartite}. We can see that the quantum entanglement between Alice and Bob disappears at the infinite curvature limit, but the quantum coherence between them  still survives. In this limit,   the subsystem Bob is fully entangled with antiBob. As a result of the monogamy of quantum entanglement, there is no entanglement between Alice and Bob. While the quantum  coherence still  exist  because only a  in the single-mode structure is required for coherence.

  From  Fig. 1(b), we can see that the coherence is a monotone-increasing function of $p$, which means that  the effect of  space curvature  can  reduce  the quantum coherence of the initial state.  It is interesting to note that  the massive field ( $\nu=0$ and  $\nu=0.2$)  preserve more quantum coherence than the massless scalar field $\nu=3/2$. That is to say, the coherence of a massive scalar  field
is more robust than a massless  field in de Sitter space. Such behavior did not appear for the quantum entanglement \cite{DS,tripartite} and steering \cite{steeringw} in the de Sitter space.

\subsection{Generating  quantum coherence between initially uncorrelated modes }

 Tracing over the mode $B$ on Eq. (\ref{All3}), we obtain the covariance matrix $\sigma_{A\bar B}$
\begin{eqnarray}\label{CMA antiB}
\sigma_{A\bar B}(s,\gamma_p)= \left(\!\!\begin{array}{cccc}
\cosh(2s)I_{1}&\frac{|\gamma_p| \sinh(2s)}{\sqrt{1-|\gamma_p|^2}}I_{1}\\
\frac{|\gamma_p|\sinh(2s)}{\sqrt{1-|\gamma_p|^2}}I_{1}&\frac{1+|\gamma_p|^2\cosh(2s)}{1-|\gamma_p|^2}I_{1}\\
\end{array}\!\!\right),
\end{eqnarray}
The corresponding symplectic value of the covariance matrix $\sigma_{A\bar B}$ is found to be  $\nu_{+}=\frac{|\gamma_p|^{2}+\cosh(2s)}{1-|\gamma_p|^{2}}$ and $\nu_{-}=1$.
with $\Delta^{(A\bar{B})}=1+\frac{(|\gamma_p|^2+\cosh(2s))^2}{(1-|\gamma_p|^2)^2}$, and $I^{A\bar{B}}_{4}=\frac{(|\gamma_p|^2+\cosh(2s))^2}{(1-|\gamma_p|^2)^2}$.
\begin{figure}[htbp]
\centering
\includegraphics[height=2.2in,width=2.8in]{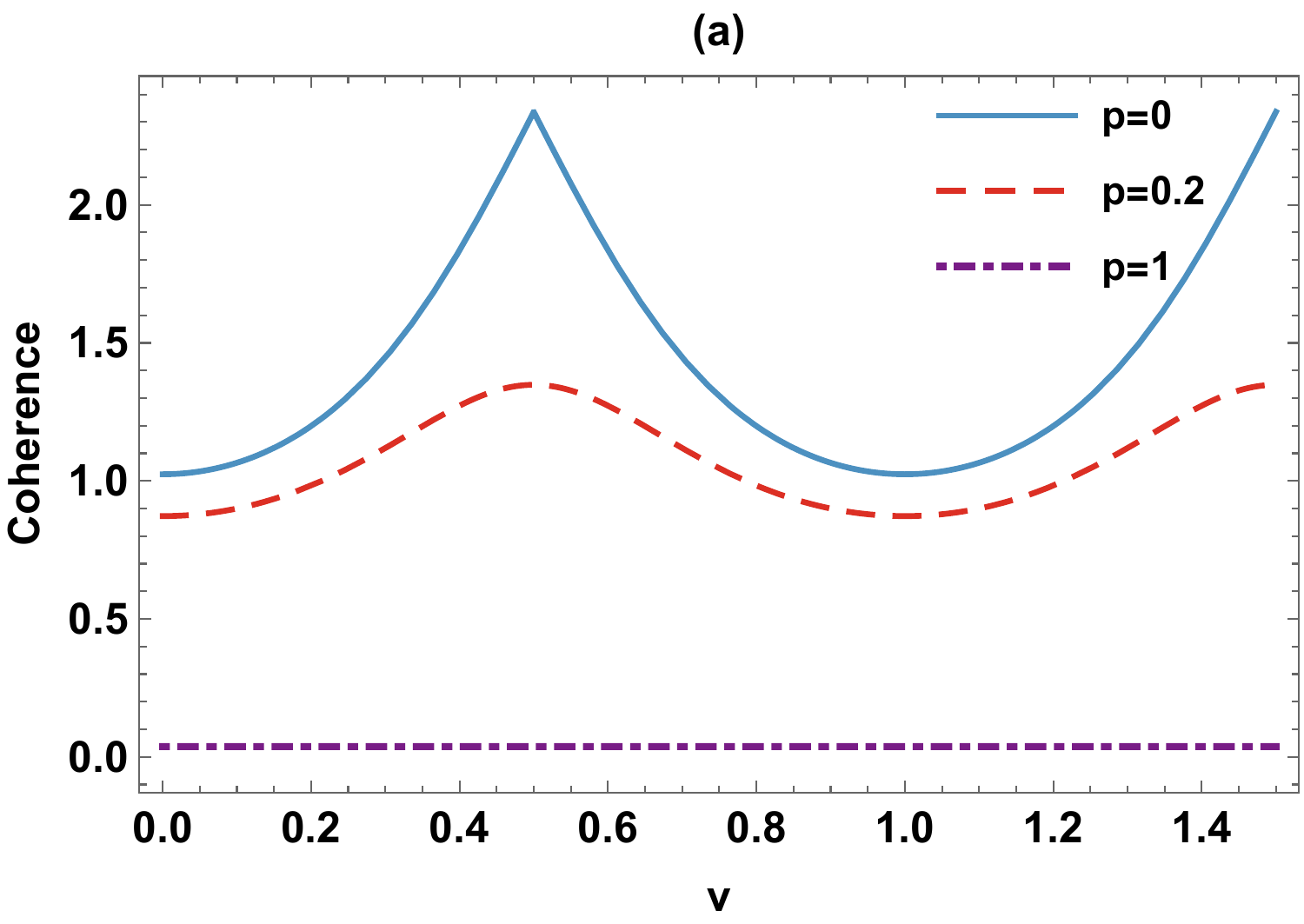}
\includegraphics[height=2.2in,width=2.8in]{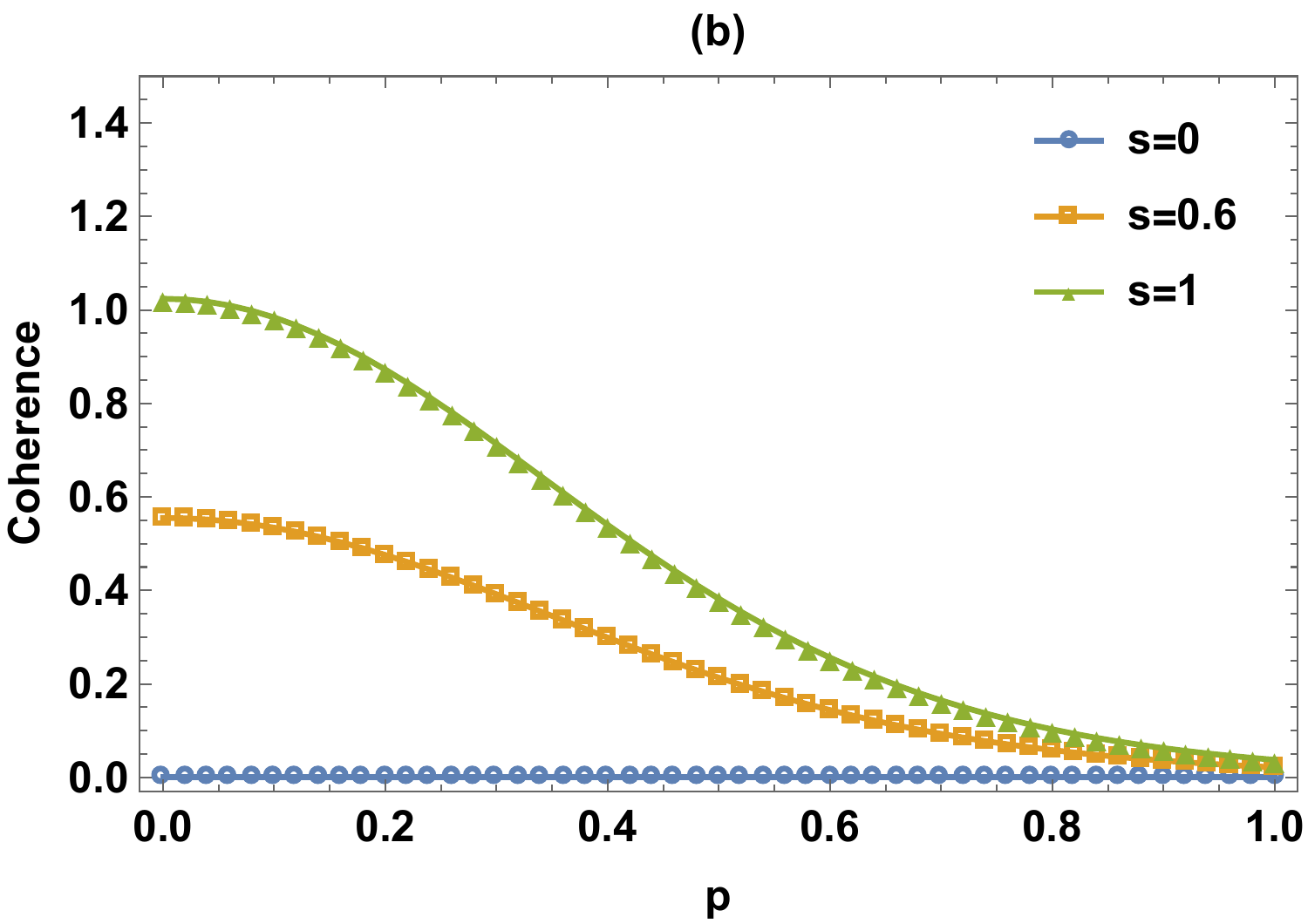}
\caption{ (Color online). (a) The Gaussian quantum  coherence between Alice and anti-Bob as a function of  mass parameters $\nu$ for different $p$. The squeezing parameter is fixed as $s=1$.  (b) The quantum  coherence as a function of the space curvature parameter $p$,  for different squeezing parameter $s=1$ (green triangle line), $s=0.6$ (orange rectangular line), $s=0$ (blue circular line). The mass parameter is fixed as  $\nu=1$.  }\label{Fig2}
\end{figure}

Fig. 2(a) shows the quantum coherence between Alice and anti-Bob concerning the mass parameters $\nu$ for a fixed initial  squeezing parameter $s=1$.
It shows that the Gaussian coherence is a periodic function versus the mass parameter $\nu$ when the space is curved ($p\neq1$).
 Like the entanglement in de Sitter space, the quantum coherence of Gaussian state $\rho_{A\bar B}$ increases   as the effect of space curvature increases. The Fig. 2(b) shows that the coherence is a monotone-decreasing function of $p$, which means the effect of space
curvature can generate quantum coherence between Alice and anti-Bob. Besides,  we can see that the larger initial state squeezing parameter, the higher Gaussian quantum coherence between Alice and anti-Bob. In the limit of flat space $p=1$, regardless of the squeezing parameter and the mass parameter, the quantum coherence about Gaussian state $\rho_{{A\bar B}}$  is  zero.

It is worth noting that  the bipartite coherence of Alice and anti-Bob is generated by the space curvature in de Sitter space, while the very same curvature also degrades the Alice-Bob coherence. We can understand this  from a  quantum information theory perspective. We know that quantum coherence can be generated by  two-mode squeezed  transformation. From Eq. (\ref{gammap3}), we can see  the Bunch-Davies vacuum for a  global  observer can be expressed as a two-mode squeezed state of the $R$ and $L$ vacua \cite{DS1,DS,DS2,discord,tripartite}.  In addition, the virtual observer anti-Bob is mapped from Bob in the chart $R$ to chart $L$.
 In other words, the curvature of the de Sitter space is in fact a two-mode squeezed transformation acts on the field modes, as has been shown in Eq. (\ref{All3}).
 Therefore, the bipartite coherence of Alice-anti-Bob is generated by the   two-mode squeezed  transformation induced by the  space curvature of the de Sitter space, while the  degradation of the Alice-Bob coherence because the initial coherence is shared among the entire tripartite system $\sigma_{A B\bar B}$.

We know that the modes $\bar B$ and $B$ are  causally disconnected because the $L$ and $R$ open charts are separated by the event horizon.  By tracing off the mode $A$, we get the covariance matrix between the observer  Bob in the $R$ region and the other observer  anti-Bob in the $L$ region, which is
\begin{eqnarray}\label{CMB antiB}
\sigma_{B\bar B}(s,\gamma_p)= \left(\!\!\begin{array}{cccc}
\frac{|\gamma_p|^2+\cosh(2s)}{1-|\gamma_p|^2}I_{1}&\frac{|\gamma_p|(\cosh(2s)+1)}{1-|\gamma_p|^2}Z_{1}\\
\frac{|\gamma_p|(\cosh(2s)+1)}{1-|\gamma_p|^2}Z_{1}&\frac{1+|\gamma_p|^2\cosh(2s)}{1-|\gamma_p|^2}I_{1}\\
\end{array}\!\!\right),
\end{eqnarray}
from which we obtain the mean occupation numbers is $\bar{n}_{B}=\frac{1}{2}(\frac{|\gamma_p|^2+\cosh(2s)}{1-|\gamma_p|^2}-1)$ and $\bar{n}_{\bar{B}}=\frac{1}{2}(\frac{1+|\gamma_p|^2\cosh(2s)}{1-|\gamma_p|^2}-1)$ of continuous variable matrix $\sigma_{B\bar{B}}$ and calculate the symplectic matrix eigenvalues  and  the quantum coherence.
\begin{figure}[htbp]
\centering
\includegraphics[height=2.2in,width=2.8in]{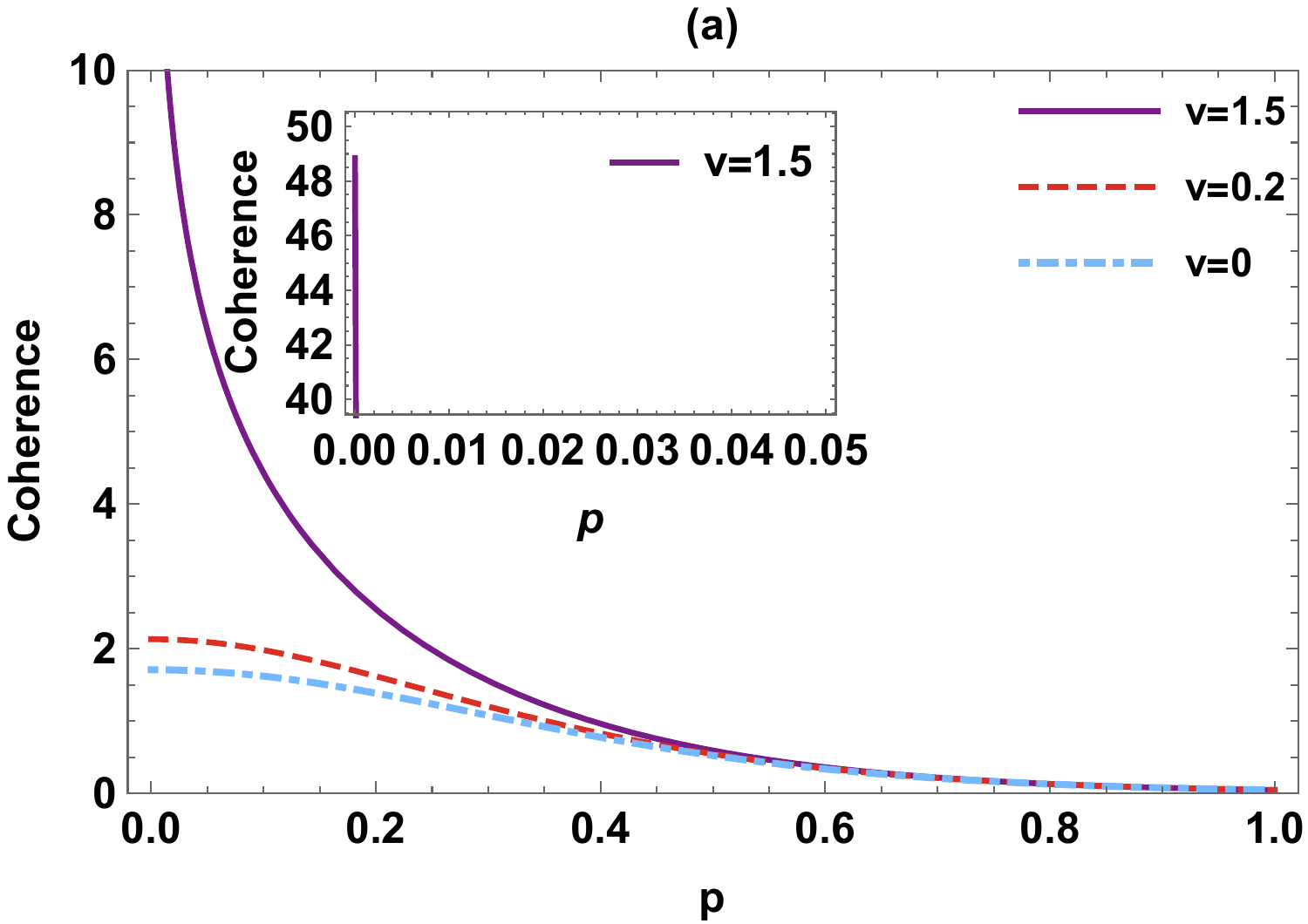}
\includegraphics[height=2.2in,width=2.8in]{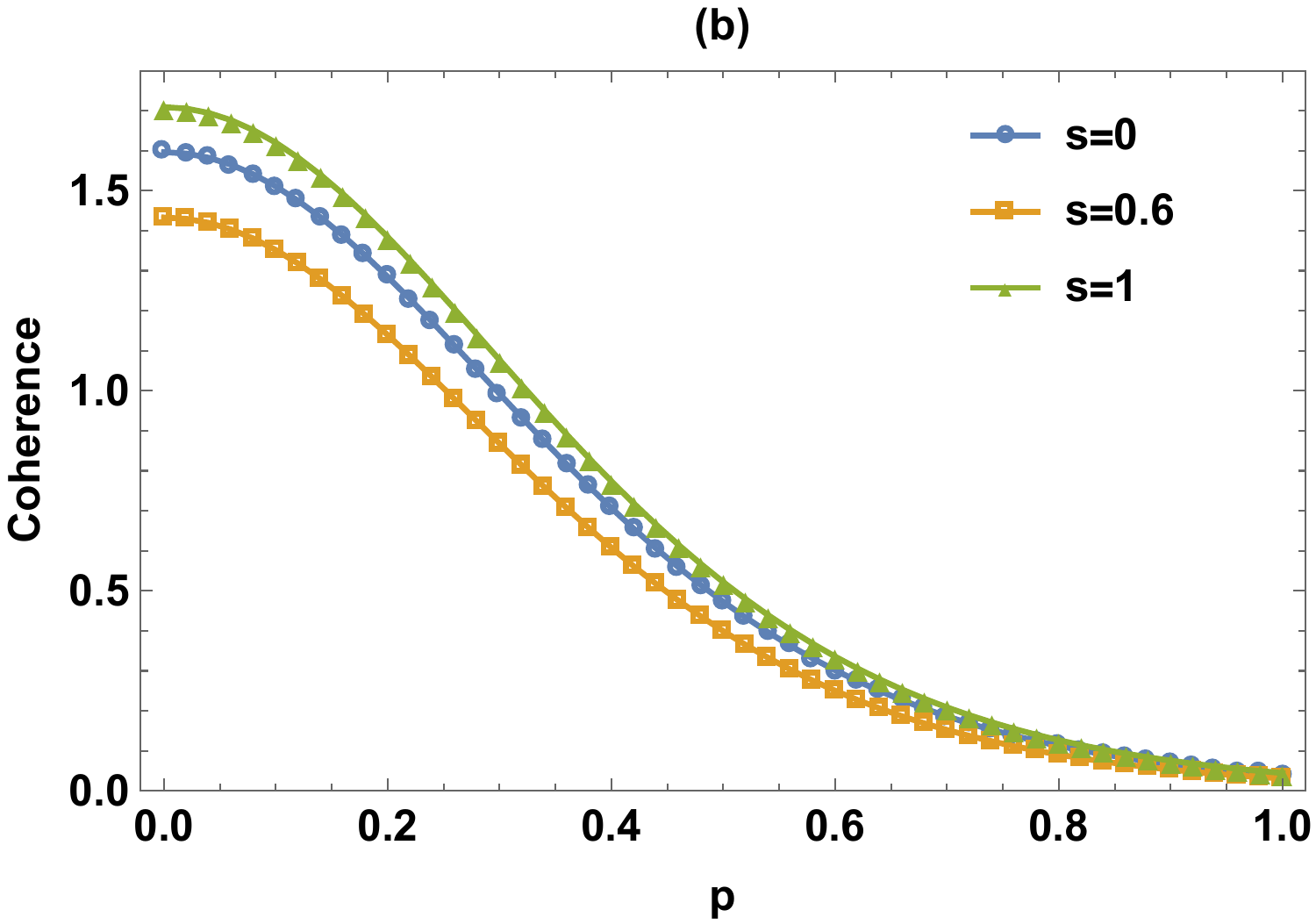}
\caption{ (color online). (a) The Gaussian quantum  coherence between Bob and anti-Bob as a function  of the parameter  $p$ . Fix the squeezing parameter $s=1$, and the blue dotted line is for $\nu=0$, the red dotted line is for $\nu=1/5$, the purple line is for $\nu=3/2$. (b) The quantum  coherence as a function of the curvature parameter $p$ for different  squeezing parameters $s=0$ (blue circular line), $s=1$ ( green triangle line ), $s=0.6$ (orange rectangular line ). Fix the mass parameter $\nu=1$. }\label{Fig3}
\end{figure}

In Fig. 3(a), it shows that the quantum coherence between modes $B$ and $\bar{B}$ increases slowly as the enhancement of the effect of space curvature  when $\nu=0$ and $\nu=0.2$.
However, for the conformally coupled
 scalar field $\nu=1/2$ and the minimally coupled massless  scalar field $\nu=3/2$, the quantum coherence for state $\rho_{B\bar{B}}$ sharply increases and approaches  to the maximum when $p\rightarrow0$. That is to say,  the  $\rho_{B\bar{B}}$  Gaussian  coherence is in fact  a sensitive indicator  for the space curvature when it in two special cases: the conformally coupled  scalar field and the minimally coupled massless scalar field.

 Fig. 3(b) also shows that the coherence is a monotone-decreasing function of $p$. The verifies that the space curvature in de Sitter space   generates the Gaussian coherence between the initially uncorrelated observers. Interestingly,
unlike the generated  Gaussian coherence between modes $A$ and $\bar{B}$,  the quantum coherence between modes $B$ and $\bar{B}$ can be very strong even the squeezing parameter is zero. The generation of Gaussian quantum  coherence between the causally disconnected regions of de Sitter space is nontrivial  because one can never communicate classically if the observers are separated by the event horizon.

\subsection{Generating  quantum coherence between three mode }
The quantum coherence among the global observer Alice and two open chart observers Bob and anti-Bob can be obtained through the covariance matrix in Eq. (\ref{All3}).
Note that $det(\sigma_{AB\bar{B}})=1$, which means the three-mode Gaussian state is pure, and the symplectic eigenvalues of this state  are equal to 1.
Thus the quantum coherence of the three-mode Gaussian state can be calculated by \cite{XU}
\begin{eqnarray}\label{threemodecohe}
C(\sigma_{AB\bar B})=\sum_{i=1}^{3}(\bar{n}_{i}+1)\log_{2}(\bar{n}_{i}+1)-\bar{n}_{i}\log_{2}\bar{n}_{i},
\end{eqnarray}
where $\bar{n}_{1}=\bar{n}_{A}$, $\bar{n}_{2}=\bar{n}_{B}$, $\bar{n}_{3}=\bar{n}_{\bar{B}}$, whose  values obtained from the covariance matrix of the three-mode Gaussian state.
\begin{figure}[htbp]
\centering
\includegraphics[height=3.0in,width=4.2in]{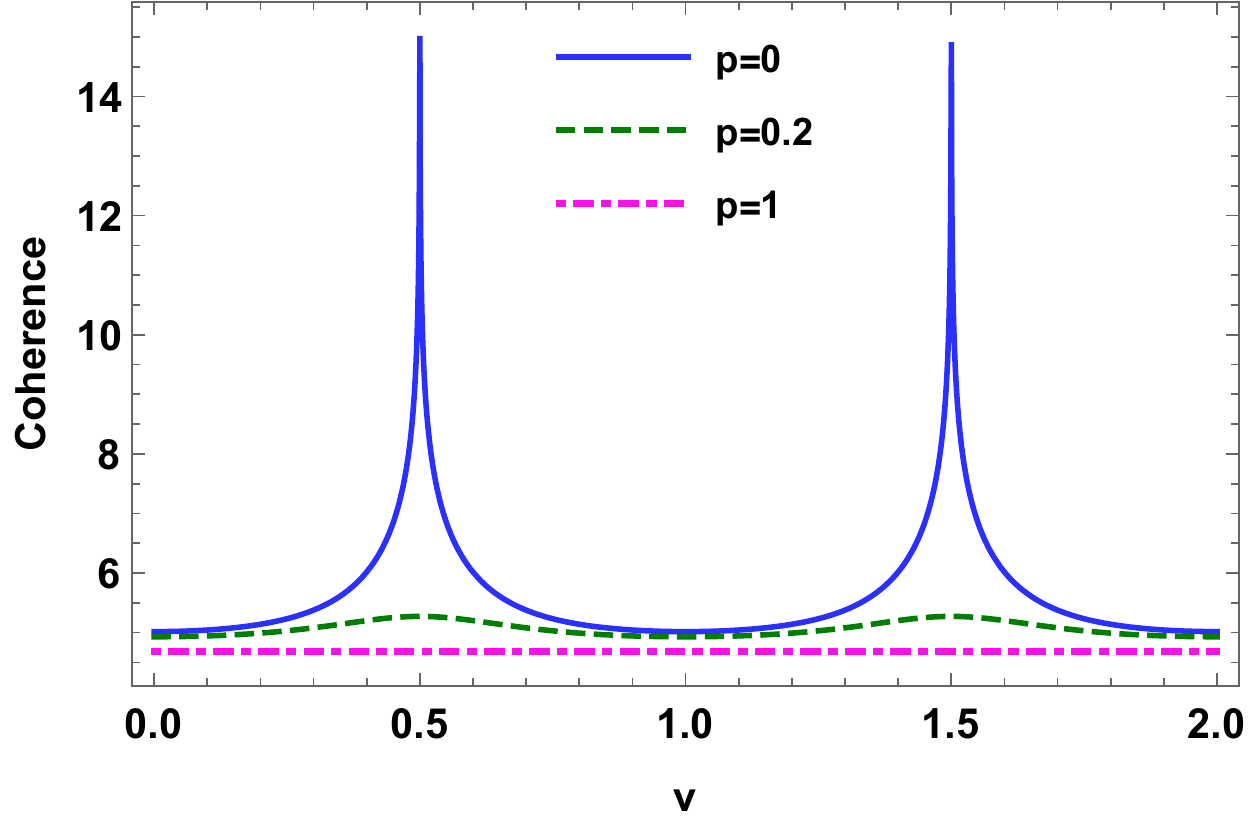}
\caption{ (color online). The three modes Gaussian quantum  coherence between Alice, Bob, and anti-Bob as functions of $\nu$ for different $p$, the squeezing parameter is fixed as $s=1$.}\label{Fig5}
\end{figure}

In Fig. 4, unlike the quantum coherence between mode $B$ and mode $\bar{B}$, the tripartite quantum coherence does not vanished in the flat space limit  $p =1$.  Besides, the  tripartite Gaussian  quantum coherence slowly changes as a function of $\nu$ for $p=0.2$.
However, the generated three-mode quantum coherence   is extremely sensitive to
 the mass parameter $\nu$ for  $p = 0$, which is very different than the generated  quantum coherence for Gaussian state $\rho_{A \bar B}$. That is to say,
the generated  three-mode Gaussian coherence is a much more sensitive indicator of space curvature in the cases of the conformally coupled
 scalar field and the minimally coupled massless  scalar field. This important  character can be in principle employed  to  design an effective detector for the space  curvature.
\section{Conclusions\label{end}}
In this work, we study  the distribution and generation  of quantum coherence for  Gaussian states  in de Sitter space.
It is found that  the curvature of the de Sitter space and  mass parameter of the field have evident effects on  the degree of coherence for all the bipartite and tripartite states.
The quantum coherence between mode $A$ and the mode $B$ decreases as the curvature effect in de Sitter space increases. However, it reaches a { \it non-zero} minimum value in the infinite curvature limit,
which is different from that of entanglement in de Sitter space.
The massive field  is found to preserve more quantum coherence than the massless scalar field for the Gaussian state $\rho_{AB}$, which did not appear for the quantum entanglement and steering. It is found that the quantum coherence is not only be distributed among  the uncorrelated modes in the
open-charts but also generates  under the curvature effect of de Sitter space.
 It shows that the quantum coherence between two causally disconnected open-charts can be very strong even if the initial state squeezing parameter is zero, which verifies the space curvature generates Gaussian quantum coherence. Unlike the generated Gaussian state $\rho_{A\bar{B}}$ coherence,  the generated three-mode and two-mode $\rho_{B\bar{B}}$ Gaussian coherence are extremely sensitive to curvature effects for conformal and massless scalar fields, which can be employed  to  design an effective detector for the space  curvature.

\begin{acknowledgments}
This work is supported by the National Natural Science Foundation
of China under Grant   No. 11875025; and Science and Technology Planning Project of Hunan Province under Grant No. 2018RS3061; and the  Natural Science Fund  of Hunan Province  under Grant No. 2018JJ1016.	

\end{acknowledgments}


\begin{thebibliography}{99}

 \bibitem{C}
T. Baumgratz, M. Cramer, and M.B. Plenio, \textit{Phys. Rev. Lett.} {\bf 2014}, \textit{113}, 140401.

\bibitem{coherencer}
A. Streltsov, G. Adesso, and M.B. Plenio, \textit{Rev. Mod. Phys.} {\bf 2017}, \textit{89}, 041003.

\bibitem{coherencer1}
M. Hu, X. Hu, J. Wang, Y. Peng, Y. Zhang, and  H. Fan, \textit{Phys. Rep.} {\bf  2018}, \textit{762}, 1-100.

\bibitem{computing1}
P.W. Shor, \textit{SIAM J. Comput.} {\bf 1997}, \textit{26}, 1484.

\bibitem{computing2}
L.K. Grover, \textit{Phys. Rev. Lett.} {\bf 1997},  \textit{79}, 325.

\bibitem{metrology1}
V. Giovannetti, S. Lloyd, and  L. Maccone, \textit{ Science} {\bf 2004},  \textit{306}, 1330.

 \bibitem{metrology2}
V. Giovannetti, S. Lloyd, and  L. Maccone, \textit{Nat. Photonics} {\bf 2011}, \textit{5}, 222.

\bibitem{biolo1}
M. B. Plenio,  and  S. F. Huelga, \textit{New J. Phys.} {\bf 2008}, \textit{10}, 113019.

\bibitem{photosynthetic}
M. J. Tao, M. Hua, N. N. Zhang, W. T. He, Q. Ai,  and F. G. Deng, \textit{Quantum Engineering} {\bf 2020}, \textit{2}, e53.

\bibitem{trans2}
M. GÃŒndoÄan, P.M. Ledingham, A. Almasi, M. Cristiani, and  H. Riedmatten, \textit{Phys. Rev. Lett.} {\bf 2012}, \textit{108}, 190504.


\bibitem{trans1}
 Y. F. Hsiao, P. J. Tsai, H. S. Chen, S. X. Lin, C. C. Hung, and C. H. Lee, \textit{Phys. Rev. Lett.} {\bf 2018}, \textit{120}, 18360.

\bibitem{Tele}
 Y. Lu, Y-C. Liu,  and Y-S. Li, \textit{Chin. Phys. B .} {\bf 2020},  \textit{29}, 060301.

\bibitem{C1}
 M. Lostaglio, D. Jennings, and T. Rudolph, \textit{ Nat. Commun.} {\bf 2015},  \textit{6}, 6383.

 \bibitem{C2}
Y. Yao, X. Xiao, and L. Ge, \textit{Phys. Rev. A} {\bf 2015}, \textit{92}, 022112.

 \bibitem{C3}
 L. Zhang, \textit{ J. Phys. A: Math. Theor} {\bf 2017}, \textit{50}, 155303.

 \bibitem{C4}
H. Zhang, B. Chen, and M. Li, \textit{Commun. Theor. Phys.} {\bf 2017}, \textit{67}, 166.

\bibitem{Grover}
G-L. Long. \textit{Physical Review A} {\bf 2001}, \textit{64}, 022307.

\bibitem{XU}
J. Xu, \textit{Phys. Rev. A.} {\bf 2016},  \textit{93}, 032111.

\bibitem{inform1}
S. L. Braunstein, and P. van Loock, \textit{ Rev. Mod. Phys.} {\bf 2005}, \textit{77}, 513.


\bibitem{relat2}
I. Fuentes-Schuller, and R.B. Mann, \textit{Phys. Rev. Lett.} {\bf 2005}, \textit{95}, 120404.

\bibitem{mode}
G. Adesso, I. Fuentes-Schuller, and M. Ericsson, \textit{ Phys.Rev.A.} {\bf 2007},  \textit{76}, 062112.

\bibitem{RQI1}
M. Aspachs, G. Adesso, and I. Fuentes-Schuller, \textit{Phys. Rev. Lett} {\bf 2010}, \textit{105}, 151301.

\bibitem{relat1}
T.G. Downes, I. Fuentes-Schuller, and T.C. Ralph, \textit{Phys. Rev. Lett.} {\bf 2011},  \textit{106},  210502.

\bibitem{RQI2}
D.J. Hosler, C. van de Bruck,  and P. Kok, \textit{Phys. Rev. A} {\bf 2012}, \textit{85}, 042312.

\bibitem{Frris}
N. Friis, A.R. Lee, K. Truong, C. Sab\'{i}n, E. Solano, G. Johansson, and I. Fuentes-Schuller, \textit{Phys. Rev. Lett.} {\bf 2013},  \textit{110},  113602.

\bibitem{RQI4}
J. Doukas, E.G. Brown, A. Dragan, and R.B. Mann, \textit{Phys. Rev. A} {\bf 2013}, \textit{87}, 012306.

\bibitem{RQI5}
D. Su, and T.C. Ralph, \textit{ Phys. Rev. D} {\bf 2014}, \textit{90}, 084022.


\bibitem{relat4}
R. Bousso, A. Shahbazi-Moghaddam, and M. Tomaevi, \textit{Phys. Rev. Lett.} {\bf 2019}, \textit{123}, 241301.

\bibitem{RQI6}
T. Liu,  J. Jing, and J. Wang, \textit{Adv. Quantum Technol.} {\bf 2018}, \textit{1}, 1800072;
J. Wang, T. Liu,  J. Jing, and S. Chen, \textit{Adv. Quantum Technol.} {\bf 2019}, \textit{2}, 1900003.

\bibitem{stee}
J. Wang, H. Cao, J. Jing J, and H. Fan, \textit{Phys. Rev. D} {\bf 2016}, \textit{93}, 125011.

\bibitem{Gyong1}
L. Gyongyosi, and I. Sandor, \textit{ Quantum Engineering} {\bf 2019}, \textit{1}, e23.


\bibitem{Gyong2}
L. Gyongyosi, \textit{ Quantum Engineering} {\bf 2020}, \textit{2}, e30.


\bibitem{DW1}
 D. Wang, W-N. Shi, R. D. Hoehn, F. Ming, W-Y. Sun, S. Kais, and L. Ye, \textit{Ann. Phys.}, {\bf 2018},  \textit{530}, 1800080.

\bibitem{DW2}
F. Ming,  D. Wang, and L. Ye, \textit{Ann. Phys.}, {\bf 2019}, \textit{531}, 1900014.

\bibitem{DW3}
F. Ming, X-K. Song, J. Ling, L. Ye, and D. Wang, \textit{Eur. Phys. J. C}, {\bf 2020}, \textit{80}, 275.

\bibitem{DS1}
 M. Sasaki, T. Tanaka, and K. Yamamoto, \textit{Phys. Rev. D} {\bf 1995},  \textit{51}, 2979-2995.

\bibitem{DS}
A. Albrecht, S. Kanno, and M. Sasaki, \textit{Phys. Rev. D} {\bf 2018}, \textit{97}, 083520.

\bibitem{DS2}
J. Maldacena, and G.L. Pimentel, \textit{ JHEP }{\bf 2013}, \textit{1302}, 038.

\bibitem{DS3}
 S. Kanno, J. Murugan, J.P. Shock, and J. Soda, \textit{JHEP} {\bf 2014}, \textit{1407}, 072.

\bibitem{DS4}
K.K. Ng, R.B. Mann, and E. Mart \~{A}­n-Mart \~{A}­­nez, \textit{Phys. Rev. D} {\bf 2018}, \textit{98}, 125005.

\bibitem{DS5}
C. Arias, F. Diaz, and P. Sundell,  \textit{Class. Quantum Grav.} {\bf 2020}, \textit{37}, 01500.

\bibitem{tripartite}
 J. Wang, C. Wen, J. Jing, and S. Chen, \textit{Phys. Lett. B} {\bf 2020}, \textit{800}, 135109.

\bibitem{DS6}
A. Matsumura. and Y. Nambu,  \textit{Phys. Rev. D} {\bf 2018}, \textit{98}, 025004.

\bibitem{discord}
S. Kanno, J.P. Shock, and J. Soda, \textit{Phys. Rev. D} {\bf 2016}, \textit{94}, 125014.

\bibitem{steeringw}
C. Wen,  J. Wang,  and J. Jing, \textit{Eur. Phys. J. C } {\bf 2020}, \textit{80}, 78.

\bibitem{brareview}
 S. L. Braunstein, and P. van Loock, \textit{Rev. Mod. Phys.} {\bf 2005},  \textit{77}, 513.

\bibitem{entropy}
A. S. Holevo, M. Sohma,  and O. Hirota, \textit{Phys. Rev. A} {\bf 1999}, \textit{59}, 1820.



\end{thebibliography}
\end{document}